\documentclass[lettersize,journal]{IEEEtran}
\usepackage{amsmath,amsfonts}
\usepackage{algorithmic}
\usepackage{algorithm}
\usepackage{array}
\usepackage[caption=false,font=normalsize,labelfont=sf,textfont=sf]{subfig}
\usepackage{textcomp}
\usepackage{stfloats}
\usepackage{multirow}
\usepackage{url}
\usepackage{verbatim}
\usepackage{graphicx}
\usepackage{booktabs}
\usepackage{cite}
\usepackage{hyperref}
\usepackage{bbding}
\usepackage{tikz}
\usepackage{comment}
\usepackage{enumitem}
\hypersetup{
  colorlinks,
  citecolor=black,
  linkcolor=black,
  urlcolor=black}

\newcommand{\tikzxmark}{%
\tikz[scale=0.23] {
    \draw[line width=0.7,line cap=round] (0,0) to [bend left=6] (1,1);
    \draw[line width=0.7,line cap=round] (0.2,0.95) to [bend right=3] (0.8,0.05);
}}
\newcommand{\tikzxmarkbold}{%
\tikz[scale=0.23] {
    \draw[line width=1.4,line cap=round] (0,0) to [bend left=6] (1,1);
    \draw[line width=1.4,line cap=round] (0.2,0.95) to [bend right=3] (0.8,0.05);
}}
\newcommand{\tikzcmark}{%
\tikz[scale=0.23] {
    \draw[line width=0.7,line cap=round] (0.25,0) to [bend left=10] (1,1);
    \draw[line width=0.8,line cap=round] (0,0.35) to [bend right=1] (0.23,0);
}}
\newcommand{\tikzcmarkbold}{%
\tikz[scale=0.23] {
    \draw[line width=1.4,line cap=round] (0.25,0) to [bend left=10] (1,1);
    \draw[line width=1.5,line cap=round] (0,0.35) to [bend right=1] (0.23,0);
}}
\newcommand{\tikzplus}{
\tikz[scale=0.23]{
    \draw[line width=0.7] (-0.5, 0) -- (0.5, 0);
    \draw[line width=0.7] (0, -0.5) -- (0, 0.5);
}
}
\newcommand{\tikzplusbold}{
\tikz[scale=0.23]{
    \draw[line width=1.2] (-0.5, -0.5) -- (0.5, -0.5);
    \draw[line width=1.2] (0, -1) -- (0, 0);
}
}
\newcommand{\tikzminus}{
\tikz[scale=0.23]{
    \draw[line width=0.7, white] (-0.5, 0) -- (0.5, 0);
    \draw[line width=0.7] (-0.5, 0.5) -- (0.5, 0.5);
}
}
\newcommand{\tikzminusbold}{
\tikz[scale=0.23]{
    \draw[line width=0.7, white] (-0.5, 0) -- (0.5, 0);
    \draw[line width=1.2] (-0.5, 0.5) -- (0.5, 0.5);
}
}

\begin{document}

\title{Cool-3D: An End-to-End Thermal-Aware Framework for Early-Phase Design Space Exploration of Microfluidic-Cooled 3DICs}

\author{Runxi~Wang,~\IEEEmembership{Graduate Student Member,~IEEE,} Ziheng Wang, Ting Lin,~\IEEEmembership{Student Member,~IEEE,} Jacob M. Raby,~\IEEEmembership{Graduate Student Member,~IEEE,} Mircea R. Stan,~\IEEEmembership{Fellow,~IEEE} and Xinfei Guo,~\IEEEmembership{Senior Member,~IEEE}        
\thanks{This work has been submitted to the IEEE for possible publication. Copyright may be transferred without notice, after which this version may no longer be accessible.}
\thanks{This work was supported in part by the National Science Foundation of China under Grant No. 62201340 and by the  Semiconductor Research Corporation (SRC) JUMP Center for Research on Intelligent
 Storage and Procesing-in-memory (CRISP).}
\thanks{Runxi Wang, Ziheng Wang, Ting Lin, Jacob M. Raby and Xinfei Guo are with the University of Michigan – Shanghai Jiao Tong University Joint Institute, Shanghai Jiao Tong University, Shanghai 200240, China (E-mails: \{wangrunxi, wangziheng1, ting\_lin, jraby8160, xinfei.guo\}@sjtu.edu.cn).}%
\thanks{Mircea R. Stan is with the Department of Electrical and Computer Engineering, University of Virginia, Charlottesville, VA 22903, USA (E-mail: mircea@virginia.edu).}
\thanks{Corresponding author: Xinfei Guo.}
}

\markboth{}%
{Shell \MakeLowercase{\textit{et al.}}: A Sample Article Using IEEEtran.cls for IEEE Journals}


\maketitle

\begin{abstract}
The rapid advancement of three-dimensional integrated circuits (3DICs) has heightened the need for early-phase design space exploration (DSE) to minimize design iterations and unexpected challenges. 
Emphasizing the pre-register-transfer level (Pre-RTL) design phase is crucial for reducing trial-and-error costs. However, 3DIC design introduces additional complexities due to thermal constraints and an expanded design space resulting from vertical stacking and various 
cooling strategies. Despite this need, existing Pre-RTL DSE tools for 3DICs remain scarce, with available solutions often lacking comprehensive design options and full customization support. To bridge this gap, we present Cool-3D, an end-to-end, thermal-aware framework for 3DIC design that integrates mainstream architectural-level simulators, including gem5, McPAT, and HotSpot 7.0, with advanced cooling models. Cool-3D enables broad and fine-grained design space exploration, built-in microfluidic cooling support for thermal analysis, and an extension interface for non-parameterizable customization, allowing designers to model and optimize 3DIC architectures with greater flexibility and accuracy. To validate the Cool-3D 
framework, we conduct three case studies demonstrating its ability to model various hardware design options and accurately capture thermal behaviors. Cool-3D serves as a foundational framework that not only facilitates comprehensive 3DIC design space exploration but also enables future innovations in 3DIC architecture, cooling strategies, and optimization techniques. The entire framework, along with the experimental data, is in the process of being released on GitHub\footnote{The GitHub link is available on \url{https://github.com/iCAS-SJTU/Cool-3D}}. 

\end{abstract}


\section{Introduction}
\begin{figure}
    \centering
    \includegraphics[width=\linewidth]{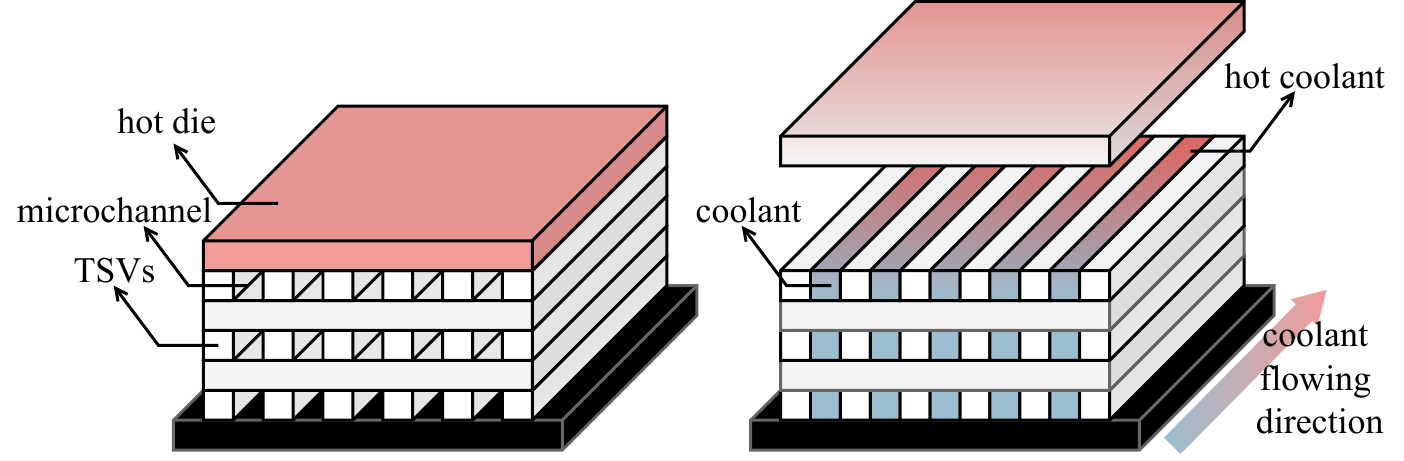}
    \caption{An illustration of working mechanism of microfluidic cooling applied in 3DICs.}
    \label{fig:microfluidic_cooling}
\end{figure}

\IEEEPARstart{T}\,he slowing down of Moore's Law has brought two-dimensional (2D) chip evolution to a 
plateau 
in energy efficiency and performance. This has led to an increased 
interest in three-dimensional integrated circuits (3DICs), where transistor density is enhanced by stacking dies vertically. 
However, 
heat dissipation and the associated thermal effects~\cite{sanipini2021thermal} become a major challenge in 3DIC development. 
In 3DICs, vertical stacking leads to heat accumulation within the dies, primarily due to the limited heat dissipation area and thermal coupling between layers. 
To address this challenge, various 3DIC-specific cooling techniques have been proposed. Table \ref{tab:comparison_cooling} lists and compares some of these techniques. Air cooling and static heatsink solutions, which are widely used for 2D chips, are not efficient enough to deal with the internally accumulated heat in a stack. Tailored for 3DICs, through-silicon-via (TSV) based cooling enhances the heat transfer across dies but its efficiency depends on dedicated placement of TSVs and still relies on other cooling medium to dissipate the heat \cite{kandlikar2011coolingoverview, salvi2021review}. 
A promising cooling method for 3DICs is microchannel or microfluidic cooling \cite{salvi2021review}. It demonstrates strong compatibility with 3D structures and has proven effective in lowering overall chip temperatures \cite{sekar20083dmicrochannel,wang20183d}. This is achieved by integrating microchannels between dies, allowing liquid coolant to circulate from an external pump as shown in Fig. \ref{fig:microfluidic_cooling}.
The challenge of this cooling method is to have well-designed microchannel patterns specific to each 3DIC to effectively carry the heat, which is a significant design step 
in the 3DIC design space. 
In addition to novel cooling strategies, optimizing the stacking configuration 
is another key approach to enhancing heat dissipation. For instance, in 3DICs handling compute-intensive workloads, placing processor core dies on the top layer, close to the heatsink, can help improve thermal management, this being another design option that needs dedicated tuning during the 3DIC designing process. 

\begin{table}[tb]
    \caption{Comparisons of 3DIC cooling methods}
    \label{tab:comparison_cooling}
    \centering
    \resizebox{\linewidth}{!}{
    \begin{tabular}{ccc}
    \toprule
         & Heat Dissipation & \multirow{2}{*}{Limits} \\
         & Efficiency & \\
    \midrule
    \midrule
        Air Cooling & - & large area needed \\
    \midrule
        Static Heat Sink & + & limited by heat transfer inside stacks \\
    \midrule
        TSV Cooling & + & special TSV arrangement needed \\
    \midrule
        Microfluidic Cooling  & ++ & special cooling pattern needed \\
    \bottomrule
    \end{tabular}
    }
\end{table}

While such 
diverse solutions to combat heat dissipation in 3DICs span multiple disciplines, from materials science, physics, thermodynamics, mathematics, to electronics, circuits and architecture, the advancement in 
each field has been separate from each other, 
partly due to the already complex and lengthy chip design cycle and missing simulation frameworks that connect the dots. As a result, thermal management, such as the cooling strategies and stacking policies mentioned above, has emerged as an additional and separate 
dimension in the 3DIC design space, introducing more design variables into an already intricate workflow. The extended design timeline and expanded design space inherent to 3D stacking further increase the cost of trial-and-error during the design and testing phases. One way to mitigate this challenge is through early-stage design space exploration (DSE) using pre-register-transfer-level (Pre-RTL) modeling and simulation, a step that becomes increasingly important and has already proven effective in 2D-based design \cite{shao2023retrospective,boroujerdian2023farsi,verma2024dso}.
Hardware modeling in Pre-RTL modeling and simulation is relatively coarse-grained, as it omits RTL-level details of the design. 
Unlike most formal design and verification flows using industry-standard Electronic Design Automation (EDA) tools, which require the transistor-level or gate-level details of a finalized design, Pre-RTL simulation operates with only a 
design concept model—such as major functional blocks and basic interconnections. Despite being somewhat abstract, 
such a high-level model can generate useful predictions of system behavior, power consumption, temperature, and other relevant metrics. 
This kind of early-phase DSE step has also been introduced 
in some commercial 3DIC EDA tools such as Synopsys 3DIC Compiler which supports early architecture exploration without RTL availability. This further highlights the significance of integrating the Pre-RTL DSE step into the 3DIC design flow. 
However, architecture design in 3DICs involve scattered design options spanning 
microarchitectural details, 
architectural hierarchy, 3D stacking configurations, and even 
cooling strategies. 
To accommodate these complexities, a Pre-RTL DSE tool framework must support a broad, hierarchical, and granularity-reconfigurable design space, ensuring comprehensive modeling of these various design options.
In 3DICs, factors such as microarchitectural details (e.g., memory hierarchy), architectural choices (e.g., instruction set architecture (ISA)), floorplanning policies, die stacking strategies, and cooling methods all influence heat generation and dissipation. Therefore, given a ``design outline'', 
it is essential to provide sufficient design options for exploring potential optimizations. Additionally, designers may need to integrate fully customized modules into large-scale designs and assess their feasibility in an agile manner. To support this, the design framework should offer an extension interface that allows seamless integration of custom modules.

Existing thermal-aware Pre-RTL DSE tools or toolchains, however, pay little focus on 3DICs or miss key elements 
that enable true DSE. For example,  HotSniper~\cite{pathania2018hotsniper} and the work in~\cite{roelke2017pre} only apply to 2DICs, however, 
they do open a way to chain mainstream specialized simulators into an end-to-end toolchain. Emerging 3DIC-oriented toolchains such as~\cite{deshwal2019moos, hankin2021hotgauge, siddhu2022comet}, have adopted a similar approach by integrating existing simulators. Nevertheless, they either lack configurability in microarchitectural details or do not account for cooling mechanisms—both of which are critical design considerations as discussed earlier. 
Additionally, these toolchains fail to support flexible customization through a uniform configuration framework, showing a lack of full tool-integration and limiting their adaptability to diverse 3DIC design needs.



When it comes to the simulators that serve as building blocks for each toolchain, they are typically designed to estimate 
a specific set of metrics. For example, performance simulators primarily model system behavior and generate switching activities, while power simulators generate power-related data. While this specialization ensures efficiency and accuracy within their respective domains, it often frustrates designers who require a metric from one tool that depends on the output of another. The primary challenges in integrating multiple simulators into a cohesive toolchain stem from mismatched input-output (I/O) interfaces and diverse runtime environments. In 3DICs, this complexity is further exacerbated by the need for thermal modeling. While existing toolchains enable 3DIC-oriented Pre-RTL DSE to some extent, there remains significant room for improvement in simulator selection and I/O compatibility. Enhancing these aspects can improve prediction accuracy and enable more fine-grained design space exploration.

To address the critical heat dissipation challenges in 3DICs, the high cost of the design process, and the growing demand for agile early-phase design exploration, this work introduces 
Cool-3D, an end-to-end thermal-aware framework. Cool-3D enables early-phase DSE with broad and fine-grained design options, advanced cooling support such as microfluidic cooling, and a user-friendly extension interface for seamless customization. Unlike toolchains that solely link existing tools like ``LEGO pieces'', Cool-3D, echoing our earlier proposal~\cite{wang2023hotlego}, enhances each integrated tool with new features, tailored interfaces, and demonstrated effectiveness in assisting early-stage decision-making. Its open-source nature further increases accessibility for the research community, fostering broader adoption and innovation.



\begin{table*}[th!]
    \centering
    \caption{Comparisons between existing DSE tools for 3DICs and this Cool-3D framework}
    \label{tab:toolchain_comp}
    \begin{tabular}{cccccc}
    \toprule 
     & MOOS \cite{deshwal2019moos} &HotGauge \cite{hankin2021hotgauge}  &CoMeT \cite{siddhu2022comet}  &  \textbf{This Work} \\
    \midrule\midrule
        Tool Integration as a Flow & \tikzxmark & \tikzcmark & \tikzcmark & \tikzcmarkbold\\
        \midrule
        Integrated Tools & - & Sniper+McPAT+3D-ICE & Sniper+McPAT+HotSpot & \textbf{gem5+McPAT+HotSpot7.0} \\
        \midrule
        Unified Input & - & \tikzxmark & \tikzxmark & \tikzcmarkbold \\
        \midrule
        3DIC Cooling & \tikzxmark & \tikzcmark & \tikzxmark & \tikzcmarkbold\\
        \midrule
        Parameterizable Customization & - & \tikzcmark & \tikzcmark & \tikzcmarkbold \\
        \midrule
        Nonparameterizable Customization & - & \tikzxmark & \tikzxmark & \tikzcmarkbold \\
        \midrule
        Target Architecture & NoC-based system & processors & processors+memory & \textbf{processors+memory} \\
    \bottomrule
    \end{tabular}
\end{table*}

\section{Background \& Related Works}\label{related_works}
For rapid and flexible design space exploration in the early design phases, most architects rely on pre-RTL simulators to identify potential design issues 
and explore better design choices. For example, design works \cite{wang2023affinity,fujiki2023mvc} employ gem5 \cite{binkert2011gem5}, ZSim \cite{sanchez2013zsim} and CACTI \cite{muralimanohar2009cacti} to model and evaluate their designs. These widely-used tools are designed as pre-RTL models and simulators. There also exist many other pre-RTL simulators that offer unique simulation scope and targets \cite{akram2019survey}. The rich ecosystem built around these 2D-based pre-RTL tools, including compiler support, significantly reduces the effort required for architects to customize and extend their simulation frameworks.


However, when shifting the focus to 3DICs, a significant gap remains between the need for early-phase design space exploration and the availability of a unified end-to-end flow. The importance of early-phase DSE stems from the increased design complexity and high costs associated with 3DICs. Unlike traditional 2D architectures, 3DICs introduce an 
expanding design space due to die stacking, making early exploration even more critical. While industry-standard EDA tools provide highly accurate and detailed analyses for completed 3DIC designs, they suffer from long runtimes and require extensive design details upfront. This raises the barrier for early-phase DSE, prolongs the design trial-and-error loop, and ultimately increases development costs.
Fortunately, research has emerged on 3DIC-supported pre-RTL thermal simulators and DSE tools. 
MOOS \cite{deshwal2019moos} is a DSE framework for network-on-chip (NoC)-based 3D manycore design. It is able to perform thermal-aware optimization, but its applicability is limited and lacks an integrated end-to-end flow.
When considering well-integrated flows, HotGauge \cite{hankin2021hotgauge} and CoMeT \cite{siddhu2022comet} are recently proposed notable toolchains. 
HotGauge \cite{hankin2021hotgauge}, making use of Sniper \cite{carlson2011sniper}, McPAT \cite{mcpat:micro}, and 3D-ICE \cite{sridhar20103dice}, presents an end-to-end flow to simulate 3DICs but mainly for processor designs. 
CoMeT \cite{siddhu2022comet}, built on Sniper \cite{carlson2011sniper}, McPAT \cite{mcpat:micro}, and HotSpot 6.0 \cite{huang2006hotspot}, supports 3DICs with both processors and memory stacking. While it offers an integrated flow, it lacks reconfigurability for self-customized designs and does not take consideration of the advanced cooling aspects in 3DICs. Consequently, its design space exploration remains coarse-grained and incomplete, particularly in the thermal dimension. Table \ref{tab:toolchain_comp} provides a clearer comparison of the key features in these existing 3DIC-oriented toolchains alongside the proposed framework Cool-3D, highlighting its superior reconfigurability and broader design space support.

\begin{figure}[tb]
    \centering
    \includegraphics[width=\linewidth]{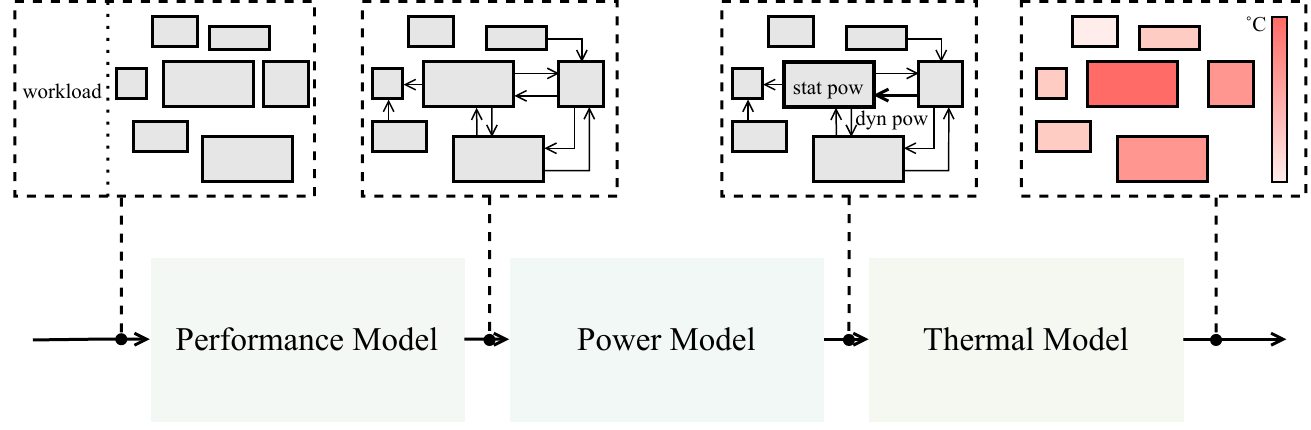}
    \caption{Overview of an end-to-end thermal simulation toolchain, comprising a performance model that generates switching activities, a power model that estimates dynamic (dyn. pow.) and static power (stat. pow.), and a thermal model that predicts heat generation.}
    \label{fig:model-flow}
\end{figure}

Based on Table \ref{tab:toolchain_comp}, the integrated tools within each toolchain vary significantly. These tools serve as foundational models and play a crucial role in shaping final design choices. Thus, selecting the appropriate tools requires a well-defined and constructive strategy. The framework requires at least three core abstract models to capture performance, power, and thermal characteristics. Fig. \ref{fig:model-flow} illustrates the interaction between these three models: given a workload and an architecture specification, the flow sequentially generates unit-level interaction data, static/dynamic power consumption, and corresponding temperature variations. This sequential modeling approach has been widely adopted and validated in 2D-based architectural simulations. For instance, HotSniper \cite{pathania2018hotsniper}, which targets 2D systems, employs Sniper, McPAT, and HotSpot 6.0 as its performance, power, and thermal models, respectively. Similarly, the 2D-based framework in \cite{roelke2017pre} integrates gem5, McPAT, and HotSpot 6.0 as its key modeling components. The success and widespread adoption of such flows in the architecture community suggest that this methodology can be extended beyond 2D chips. However, selecting the appropriate tools for each model involves navigating a diverse landscape of available simulators. For example, gem5 can be used as an alternative to Sniper as a performance model due to its unique features. The choice ultimately depends on the targeted architecture and the specific capabilities required for simulation and DSE. The following sections will analyze the differences among mainstream computer architecture simulators used for performance, power, and thermal modeling and justify the selection criteria for the proposed Cool-3D framework. This selection process constitutes a key component of our proposed design flow.

\subsection{Selection of Performance Model}
Gem5 \cite{binkert2011gem5} and Sniper \cite{carlson2011sniper} are two widely used performance simulators in computer architecture research. As shown in Table \ref{tab:perf_comp}, their primary distinction lies in their modeling methodologies. Sniper employs an interval simulation approach, which reduces modeling time, whereas gem5, as an event-driven simulator, tracks events cycle by cycle \cite{akram2019survey}. This makes gem5 more accurate than Sniper but at the expense of longer simulation times. As the starting point of the 3DIC simulation toolchain, the performance model must provide detailed output statistics to enable subsequent stages to capture richer information, leading to more accurate thermal predictions. Thus, finer simulation granularity is a key criterion for selecting the performance model. Compared to Sniper, gem5 offers more detailed statistics for individual units. Furthermore, for architects requiring the integration of customized modules, gem5 provides a more flexible and fine-grained extension interface. Considering both accuracy and extensibility, gem5 has been selected as the most suitable choice for our framework.

\begin{table}[tb]
    \caption{Feature Comparisons between gem5 and Sniper}
    \label{tab:perf_comp}
    \centering
    \resizebox{\linewidth}{!}{
    \begin{tabular}{lcccc}
    \toprule
        & Accuracy & Running      & Simulation             & Customization \\
        &          & Time         & Granularity            & Support \\
    \midrule
    gem5 \cite{binkert2011gem5} 
        & \tikzplusbold & \tikzplus & \tikzplusbold & \tikzplusbold\tikzplusbold \\
    Sniper \cite{carlson2011sniper} 
        & \tikzminus & \tikzminus & \tikzminus & \tikzplus \\
    Needs of This Work 
        & \tikzplusbold & \tikzminusbold & \tikzplusbold & \tikzplusbold\tikzplusbold \\
    \bottomrule
    \end{tabular}}
\end{table}
\subsection{Selection of Power Model}
For power modeling, McPAT \cite{mcpat:micro} has been widely used in computer architecture research for its power prediction capabilities and configurable support for multicore and manycore processors. It takes microarchitectural details and performance statistics from a performance model and outputs estimates for dynamic, static, and short-circuit power. CACTI \cite{muralimanohar2009cacti}, another popular power model, specializes in memory and cache modeling. A later version CACTI-3DD \cite{chen2012cacti3dd} extends support to 3D-stacked memory, making it particularly relevant for 3DICs. Like McPAT, CACTI derives power and energy estimates from performance model data. Typically, McPAT models processor cores, while CACTI handles memory subsystems, allowing for a comprehensive power analysis of both compute and memory components. 
Several McPAT variants extend upon the original, as summarized in Table~\ref{tab:mcpatvs}. McPAT-Calib \cite{zhai2022mcpat} integrates microarchitecture simulation, power modeling, and machine learning (ML)-based calibration to refine McPAT’s power estimates. However, its reliance on architecture-specific training data limits its generality for broader design exploration. New features introduced by McPAT-Monolithic~\cite{guler2020mcpat} require FinFET process node libraries and gate-level simulation, diverging from the objectives of Pre-RTL modeling. Given these considerations, McPAT combined with CACTI-3DD offers the most suitable power modeling solution for our 
work.
\begin{table}[tb]
    \caption{Feature Comparisons of McPAT, McPAT-calib, and McPAT-Monolithic}
    \label{tab:mcpatvs}
    \centering
    \resizebox{\linewidth}{!}{
    \begin{tabular}{lcccc}
    \toprule
    & ML & General & Pre-RTL  & Simulation \\
    & Assisted & Purpose & Compatibility & Level\\
    \midrule
    McPAT \cite{mcpat:micro} & \tikzxmarkbold & \tikzcmarkbold & \tikzcmarkbold & \textbf{Arch-level} \\
    McPAT-Calib \cite{zhai2022mcpat} & \tikzcmark & \tikzxmark & \tikzcmark & Arch-level \\
    McPAT-Monolithic \cite{guler2020mcpat} & \tikzxmark & \tikzcmark & \tikzxmark & Circuit-level \\
    Needs of This work & \tikzxmarkbold & \tikzcmarkbold & \tikzcmarkbold & \textbf{Arch-level} \\
    \bottomrule
    \end{tabular}}
\end{table}

\subsection{Selection of Thermal Model}
There exists a wide range of thermal simulations \cite{sultan2019survey}, Hotspot 6.0 \cite{huang2006hotspot} has been a widely used tool for modeling temperature distributions based on power traces. It leverages the analogy between electrical circuits and heat conduction to efficiently solve one-dimensional steady-state heat conduction problems. Its accuracy has been validated against real chips \cite{zhang2015hotspot}. The latest version, HotSpot 7.0, introduces support for microfluidic cooling \cite{han2021hotspot,han20222}, which is essential and unique for introducing advanced cooling features to the designer. Among alternative thermal models, 3D-ICE \cite{sridhar20103dice} also supports microfluidic cooling and has been experimentally validated \cite{sridhar20133dice}. However, HotSpot 7.0 is preferred for its broader applicability in full-stack thermal-aware architecture design and its well-established compatibility with McPAT. There exist several toolchains that integrate 3D-ICE, such as \cite{huang2024evaluation, hankin2021hotgauge}, but the full tool integration is still under development. Other recent thermal simulators, such as the Manchester Thermal Analyzer (MTA) \cite{ladenheim2018mta} and ARTSim \cite{safari2023artsim}, also model thermal effects in 3DICs. However, MTA lacks microfluidic cooling support, and neither has been integrated into a stable 3DIC-oriented toolchain. 
There are also ML-based thermal predictors introduced in recent works such as FaStTherm \cite{zhu2024fasttherm} and the work in \cite{chhabria2021thermal}. They feature quick thermal prediction, but their reliance 
on the dataset and training process is out of scope for this work because we focus on the generality of our approach. 
Novel thermal models such as \cite{xu2025wafer} targeting wafer-scale heterogeneous integration (WSHI) are also out of our scope for its different applicable integration technology. 
Table \ref{tab:thermal_simulators} summarizes the key differences among these thermal models and highlights our selection. This work selects HotSpot for its proven compatibility with McPAT, a pairing extensively validated in existing toolchains, and specifically adopts version 7.0 to leverage its microfluidic cooling capabilities.

\begin{table}[tb]
    \caption{Features Comparisons of 3D-ICE, ARTSim, MTA, and HotSpot 7.0}
    \label{tab:thermal_simulators}
    \centering
    \resizebox{\linewidth}{!}{
    \begin{tabular}{lccccc}
    \toprule
    & Microfluidic  & 3DIC & Verified on & Compatibility & ML\\
    & Cooling Support & Compatible & Real Chips & with McPAT & Assisted\\
    \midrule
    3D-ICE\cite{sridhar20103dice, terraneo20223dice}        & \tikzcmark & \tikzcmark & \tikzcmark & \tikzplus & \tikzxmark\\
    ARTSim\cite{safari2023artsim}        & \tikzcmark & \tikzcmark & \tikzxmark & $\mathbf{-}$ & \tikzxmark \\
    MTA\cite{ladenheim2018mta}           & \tikzxmark    & \tikzcmark & \tikzxmark & $\mathbf{-}$ & \tikzxmark \\
    HotSpot 7.0\cite{huang2006hotspot, zhang2015hotspot, han2021hotspot}   & \tikzcmarkbold & \tikzcmarkbold & \tikzcmarkbold & \tikzplusbold\tikzplusbold & \tikzxmarkbold \\
    FaStTherm \cite{zhu2024fasttherm} & \tikzxmark & \tikzxmark & \tikzxmark & \tikzplus & \tikzcmark \\
    Work in \cite{chhabria2021thermal} & \tikzxmark & \tikzxmark & \tikzxmark & \tikzminus & \tikzcmark \\
    Needs of This Work     & \tikzcmarkbold & \tikzcmarkbold & \tikzcmarkbold & \tikzplusbold\tikzplusbold & \tikzxmarkbold\\
    \bottomrule
    \end{tabular}}
\end{table}

Given the limitations of existing 3DIC-oriented toolchains and our careful selection of models for performance, power, and thermal analysis, we propose Cool-3D, an end-to-end framework for thermal-aware DSE in the early design phase. Our main contributions are as follows:
\begin{enumerate}[leftmargin=0.45cm]
\item Cool-3D is a thermal-aware 3DIC simulation framework with full tool integration as the back-end and a unified input format as the front-end for rapid Pre-RTL simulation and efficient design space exploration.
\item The Cool-3D framework supports extensive configurability, covering microarchitectural details and 3D stacking configurations to accommodate diverse design needs.
\item Cool-3D features microfluidic cooling support with a floorplan generator and a microfluidic cooling strategy generator, allowing designers to fully customize cooling configurations.
\item This framework offers an extension interface for integrating fully customized modules into the simulated architecture.
\item The framework will be fully open-sourced and continuously improved to support future research and development.
\end{enumerate}

\section{The Proposed End-to-End Thermal-Aware Framework Cool-3D}
This section presents the details of the proposed Cool-3D framework, starting with a high-level overview of its integration 
and functionality, followed by a detailed breakdown of each key feature.

\subsection{Framework Overview}
\begin{figure*}
    \centering
    \includegraphics[width=\linewidth]{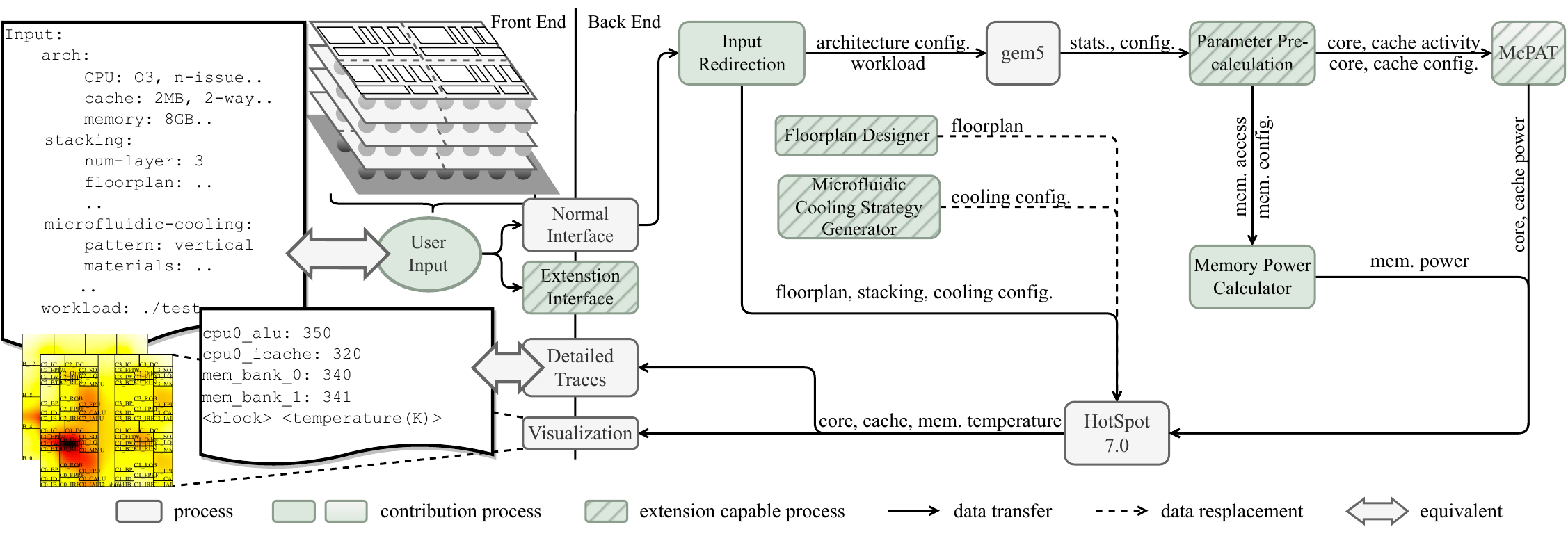}
    \caption{An overview of the proposed Cool-3D flow, composing a front-end interface and a back-end toolchain.}
    \label{fig:overall-flow}
\end{figure*}

\begin{figure}
    \centering
    \includegraphics[width=\linewidth]{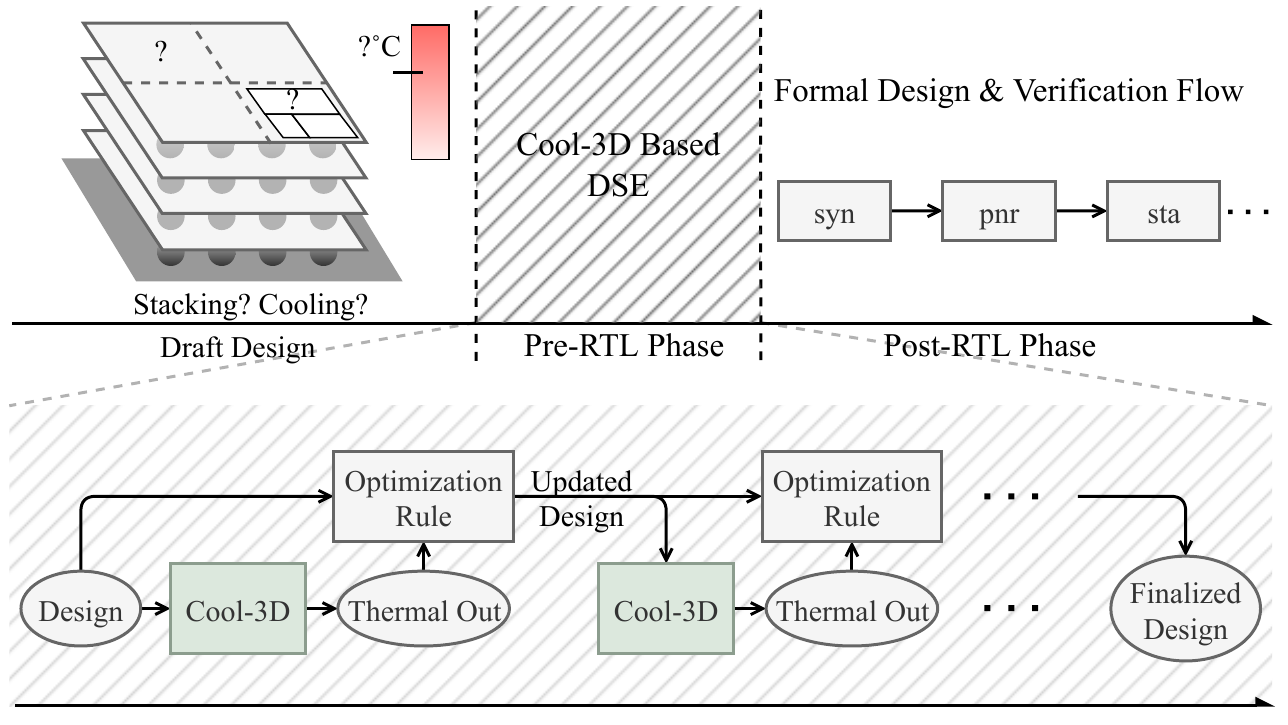}
    \caption{Illustration of the role Cool-3D plays in a typical design flow for 3DICs.}
    \label{fig:cool3d_in_design_flow}
\end{figure}
As shown in Fig. \ref{fig:overall-flow}, the Cool-3D framework consists of a user-friendly front end and a back-end toolchain, forming an agile flow that translates high-level design abstractions into concrete thermal traces and visualized results such as the heat map. The interaction and connection logic are also illustrated, with the key contributions of this work highlighted in green boxes.
The front end mainly handles user inputs and toolchain outputs,  abstracting the underlying workflow to reduce the learning curve for using Cool-3D and facilitating the generation of expected results.
The back-end toolchain integrates three carefully selected simulators as discussed earlier in \ref{related_works}.
Gem5 \cite{binkert2011gem5}, as the performance model, consumes workloads, which are the program executables, and architecture/microarchitectural configurations as inputs, and outputs the statistics representing interactions between units. 
McPAT \cite{mcpat:micro}, used in the subsequent step for power modeling, utilizes processed gem5 outputs to calculate core and cache power values. Following a similar manner, the memory power calculator generates the power prediction for 3D-stacked memory. 
As the final step, HotSpot 7.0 \cite{han2021hotspot}, for thermal modeling, takes power data as well as the floorplaning, stacking, and cooling information to generate thermal prediction. The fusion of these tools to achieve the targeted functionality of Cool-3D is elaborated in \ref{tool_fusion_sec}.

Furthermore, Fig. \ref{fig:cool3d_in_design_flow} illustrates the essential role Cool-3D plays within a 3DIC design cycle.  Given a rough 3DIC design incorporating thermal-aware optimization, designers can use Cool-3D to evaluate their effectiveness and obtain quick feedback. The 
design outline is abstracted as input to the Cool-3D front end, enabling an iterative design space exploration (DSE) process. Designers can define their own optimization rules and interact with the flow, iterating through multiple refinements until a satisfactory thermal profile is achieved. Once optimized, the finalized design proceeds to formal design and verification in a ``post-RTL'' EDA toolchain.


\subsection{Tool Fusion for Rapid DSE}\label{tool_fusion_sec}
Tool fusion is a critical step in constructing Cool-3D, as each simulator in the toolchain contributes to only a portion of the overall result prediction. One of the most challenging obstacles in using existing Pre-RTL simulators is the inconsistency in working environments, diverse input formats, and the lack of a seamless transition interface between tools. These issues result in a steep learning curve and a high likelihood of errors during experimentation. To address this, Cool-3D provides a fully integrated workflow, eliminating the need for designers to manually customize their own flow for thermal-aware 3DIC development.
The first step in this integration is unifying input configurations across the three simulators. Instead of requiring designers to handle individual setups, Cool-3D introduces an input redirection process that translates simplified user inputs from the front end into structured configurations for each simulator, as shown in Fig. \ref{fig:overall-flow}. Through this process, gem5 receives workload and architectural/microarchitectural details, while HotSpot 7.0 receives floorplan and stacking information for thermal analysis. By automating these translations, Cool-3D reduces complexity, minimizes errors, and accelerates the design process for thermal-aware 3DIC development.

The next step in the tool fusion process is I/O matching between gem5 and McPAT, achieved through the parameter pre-calculation process shown in Fig. \ref{fig:overall-flow}. While existing translation scripts \cite{parserscript1, parserscript2, parserscript3} provide some level of compatibility between these tools, they suffer from critical limitations in handling complex architectural configurations. Issues such as version incompatibility, incomplete and inaccurate parameter mapping, and poor error resilience often force architects to rely on outdated simulation tools and manually reconcile discrepancies between performance and power estimation. To overcome these challenges, Cool-3D introduces an adaptive parameter translation framework that modernizes compatibility, enhances parameter mapping, and improves error handling, ensuring a seamless and automated transition from gem5’s performance outputs to McPAT’s power estimation. This adaptive parameter translation framework extends the existing translation-based approach to map gem5 outputs into McPAT-compatible parameters. 

For power estimation of cores and caches, Cool-3D utilized an enhanced version of McPAT, which we have modified specifically for our framework, as highlighted in Fig. \ref{fig:overall-flow}. In our modifications, we focused on two key areas: efficiency in simulation and a cleaner output interface. In the original version of McPAT, the initialization phase and power simulation phase are coupled together. This coupling can be time-consuming, particularly for certain configurations, leading to inefficiency in the consecutive simulation timestamps as a significant portion of the time is spent on redundant re-initialization. This inefficiency remains a common issue in many existing toolchains that use McPAT. To address this, we refactored McPAT to decouple the initialization and power simulation phases, thereby allowing for more efficient trace simulation. This change significantly reduces the time spent in initialization and speeds up the overall transient simulation process. Furthermore, we enhanced the output interface by adding a power trace feature that aligns with the needs of common thermal simulators. Instead of the complex format conversion typically required by external scripts, the modified output interface directly provides the summation of power per component. This streamlined interface reduces communication complexity within the toolchain and enhances the overall efficiency of the power estimation process.


While McPAT provides detailed power predictions, it does not specifically support 3D-stacked memory. To address this gap and improve modeling accuracy, we integrate CACTI-3DD \cite{chen2012cacti3dd} as an additional step in the power estimation process, particularly for generating power consumption references for the 3D-stacked memory. Compared to previous works, such as CoMeT \cite{siddhu2022comet}, which also use CACTI-3DD for memory power generation, Cool-3D offers a more complete integration. By incorporating CACTI-3DD directly into the toolchain, Cool-3D allows users to dynamically adjust the configuration of CACTI-3DD at the front end. This dynamic configuration enables the tool to generate memory power data on the fly, providing real-time power estimations for the 3D memory stack during simulation.



The final step in the back-end toolchain involves running thermal simulations using HotSpot 7.0, which takes power traces generated by McPAT and the memory power calculator as inputs. It is important to note that HotSpot 7.0 also requires several additional configurations, such as floorplaning, stacking configuration, cooling, and material parameters. These configurations are provided by the input redirection process, as shown in Fig. \ref{fig:overall-flow}. However, Cool-3D offers greater flexibility by allowing designers to customize their floorplan and microfluidic cooling strategies. The tool enables users to create and implement their own floorplan designs and cooling strategies, which can replace the default ones provided by the toolchain. These customized configurations help optimize the thermal performance of the design. The details of the floorplan designer and the microfluidic cooling strategy generator will be discussed further in \ref{microfluidic}, highlighting how these tools assist in fine-tuning thermal management for 3D IC designs.

\subsection{Microfluidic Cooling Support with Configuration}\label{microfluidic}
\begin{figure}[tb]
    \centering
    \includegraphics[width=\linewidth]{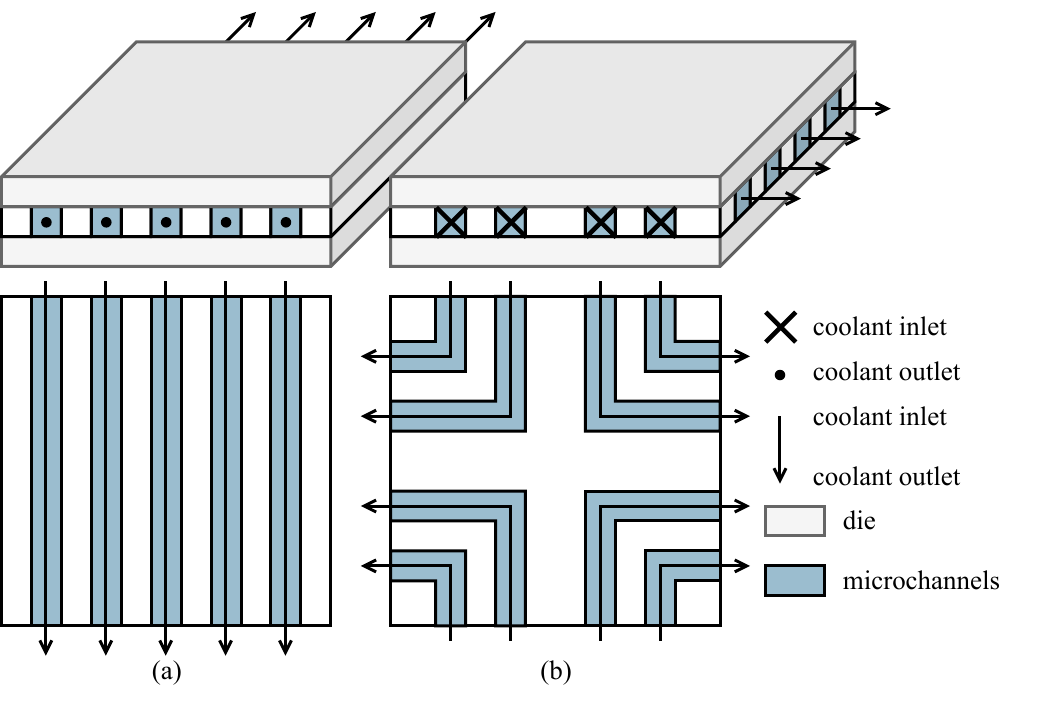}
    \caption{An illustration of mainstream microfluidic cooling patterns, (a) vertically aligned pattern, (b) 90-degree bent pattern with two pairs of inlets and outlets.}
    \label{fig:cooling_pattern}
\end{figure}
\begin{figure*}[tb]
    \centering
    \includegraphics[width=\linewidth]{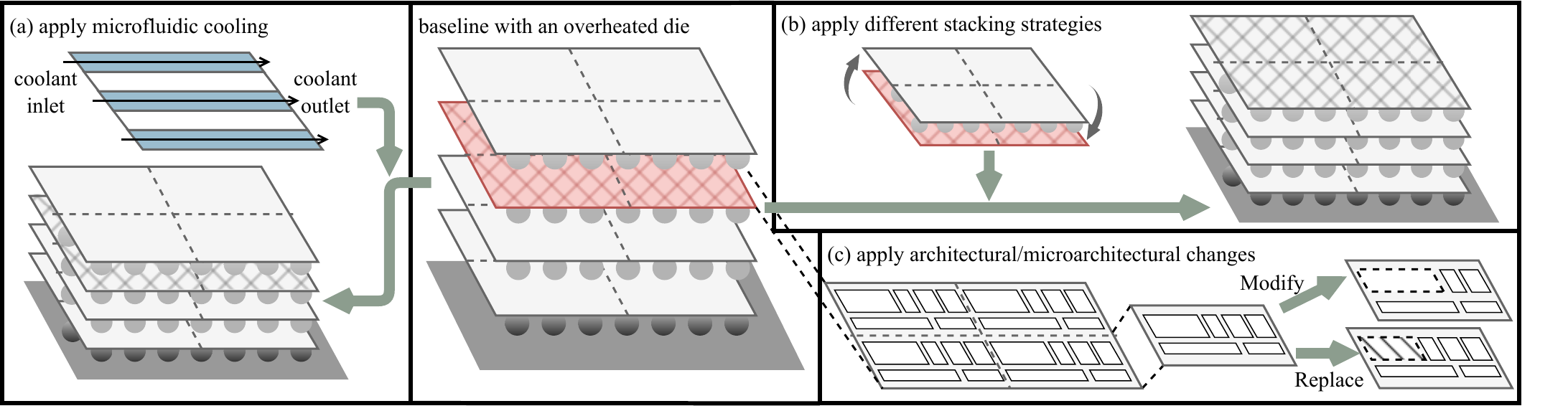}
    \caption{A demonstration of supported design space options for an overheated baseline 3DIC design in Cool-3D. (a) application of configurable microfluidic cooling methods; (b) alternation of die stacking; (c) architectural or microarchitectural design changes.}
    \label{fig:dse_space}
\end{figure*}
Cooling effects play a crucial role in shaping the thermal behavior of the input 3DIC design within the thermal model. As discussed in \ref{related_works}, HotSpot 7.0 was selected as the thermal model for Cool-3D due to its built-in support for 3D-stacked chips, its integrated microfluidic cooling mechanism, and its strong compatibility with McPAT. Additionally, HotSpot 7.0 provides an intuitive user input interface for setting microfluidic cooling channel layers, allowing for a compact yet effective abstraction of complex microfluidic cooling configurations. This feature simplifies the integration of advanced cooling strategies, enhancing the accuracy and usability of the thermal simulations within Cool-3D. 
However, most existing simulators or toolchains rely on users' full mastery of each design option in a large 3DIC system and require users to provide complex input configurations. This raises the bar to entry, particularly for designers focusing on only a subset of design options. For example, ArchFP \cite{faust2012archfp} is a floorplan designer intended to work with HotSpot, generating the required input floorplan files. However, Cool-3D does not integrate this tool due to its lack of an input interface and the requirement for users to modify source code and recompile for each customization. To enhance usability and facilitate DSE by allowing users to define a range of design options quickly, Cool-3D introduces a built-in floorplan designer and a microfluidic cooling strategy generator. These tools enable the autonomous generation of customized floorplans and cooling configurations, leveraging HotSpot 7.0’s features while ensuring ease of use and flexibility.

The floorplan designer in Cool-3D has three modes corresponding to three different customization levels. 
First, for users who do not specify a standard floorplan and only require a coarse-grained thermal result, the floorplan designer will generate a default version of the floorplan using the templates already integrated in this framework. The generation process in this mode will take the user input, such as the die area, the number of cores, and the number of memory banks into account to ensure the basic matching of hardware configurations. 
Second, to have more details in the thermal outputs and be more accurate in locating the hotspot in the chip, users have the option to automatically generate the floorplan from the McPAT output. Making use of the reference area data from McPAT output, the program can generate an initial version of the floorplan with which the user can run directly with Cool-3D. Users can adjust and iteratively update the floorplan later according to design needs. 
Finally, to include more customization freedom for the floorplan, along with the option to manually input a well-formatted floorplan file, Cool-3D integrates a floorplan designer graphical user interface (GUI). This allows users to easily experiment with different unit placement options without manually calculating the dimensions and coordinates of each unit as the program automatically generates the correctly formatted floorplan from what was designed with the GUI.
The three modes of the floorplan designer offer three different levels of configuration granularity so that users exploring different design options can quickly get started with Cool-3D without the efforts of fully preparing all the needed input files. 

Besides the floorplan, configuring microfluidic cooling patterns can also be challenging for designers, yet it is critical for consideration as it impacts the maximum thermal design power (TDP) a chip layer is able to consume. While manually defining microchannel configurations within the HotSpot 7.0 tool is feasible, an automated process enhances the adaptability of simulations. The microchannel geometries must align with the simulation resolution set in Cool-3D to ensure effective cooling coverage across the entire die. And the actual microchannel placement, as a critical design option in 3DIC designing, should be 
designed such that the coolant can enhance the heat dissipation. 
Some microfluidic cooling patterns are shown in Fig. \ref{fig:cooling_pattern}. Fig. \ref{fig:cooling_pattern} (a) depicts one of the most commonly used cases, vertically aligning microchannels between two dies. A similar pattern to this vertical one is the horizontal version. These two patterns only have one inlet and one outlet on opposite sides. A more complex case can have two or more for both. Shown in Fig. \ref{fig:cooling_pattern} (b), the microchannels are 90-degree bent with two inlets attached on opposite sides and two outlets attached on the other pair of opposite sides. This pattern is used in \cite{brunschwiler2009validation, sridhar20133dice} showing great effectiveness compared with the previously mentioned ones. All the discussed patterns are supported by the proposed microfluidic cooling strategy generator, which can provide verified cooling patterns. 
This generator allows users to specify a desired cooling pattern, automatically generating a HotSpot-compatible configuration with a simulation resolution derived from the front-end input, streamlining the evaluation of microfluidic cooling strategies.

\subsection{Hyper-Dimensional Configurable Design Space}\label{dse}
3DIC design introduces a ``hyper-dimensional'' design space compared to the traditional 2D designs. Even after finalizing the microarchitectural and architectural aspects, additional design space dimensions emerge, including  floorplanning, die stacking strategies, and cooling configurations. Each of these dimensions present unique optimization opportunities and expand the design space significantly.

Existing toolchains either lack the fine-granularity of design modeling, such as microarchitectural details, or fail to capture the 3DIC-specific cooling methods. Cool-3D on the other hand, designed to bridge this gap, supports configurations across all design space dimensions while preserving detailed outputs at the microarchitectural level. Fig. \ref{fig:dse_space}  illustrates examples using the design options supported by Cool-3D across three distinct design space dimensions, showcasing its flexibility in exploring various 3DIC configurations.
For a 3DIC design that can be potentially overheated in specific layers, one straightforward approach is to apply ``stronger'' cooling techniques directly to the affected layer. As indicated in Fig. \ref{fig:dse_space} (a), applying a microfluidic cooling layer near the overheated layer will help to reduce the overall temperature. Designers with Cool-3D can also customize the microchannel geometries to find the best way to dissipate the heat. 
A more effective approach involves optimizing stacking policies. Customized stacking rules can be seamlessly integrated into the feedback loop shown in Fig. \ref{fig:cool3d_in_design_flow} to determine the optimal organization of the entire stack. Fig. \ref{fig:dse_space} (b) presents a simplified example of this implementation, demonstrating how strategic stacking can enhance thermal management and overall performance.
For more fine-grained optimization, Cool-3D allows direct configuration of architectural and microarchitectural details through its input interface. Designers can easily modify a CPU core template or adjust internal parameters for specific modules, as demonstrated in Fig. \ref{fig:dse_space} (c). Overall, the built-in design options in Cool-3D provide designers with extensive flexibility to effectively optimize their designs with multiple design knobs.

\subsection{Extension Interface for Non-Parameterizable Customization}
\begin{figure}[tb]
    \centering
\includegraphics[width=\linewidth]{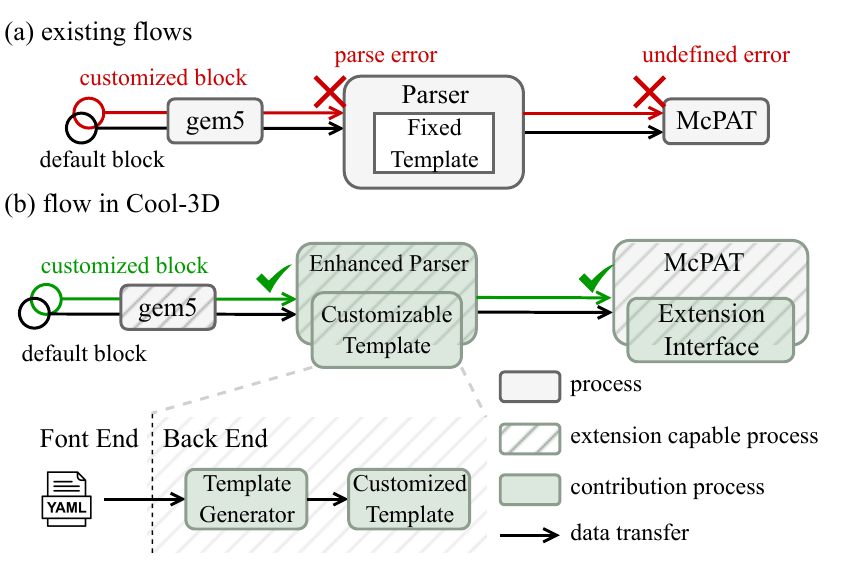}
    \caption{Demonstration of working mechanism for the nonparameterizable customization support in Cool-3D.}
    \label{fig:extension_interface}
\end{figure}

While the design space described in \ref{dse} offers a wide range of design options for designers, unconventional design ideas may arise and are needed to be modeled during the design and simulation process. Such design ideas typically can be categorized into two types: parameterizable customization, which can be configured using the front-end input interface, and non-parameterizable customization, where designers define entirely new modules and behaviors. Existing 3DIC-oriented toolchains do not support non-parameterizable customization, but Cool-3D introduces this feature to help designers seamlessly integrate self-defined modules and observe their thermal behaviors. As shown in Fig. \ref{fig:extension_interface}, the implementation of non-parameterizable customization consists of two key elements, the design parameter pre-calculation process before the power model and an extended version of McPAT with an extension interface.


The design parameter pre-calculation process involves two parallel tasks, fetching memory access traces and memory configurations for the memory power calculator, and pre-calculating in-core activity and configurations for McPAT inputs. Since non-parameterizable customization of memory can be achieved through modifications in the front-end input, our primary focus is on customizing cores or caches, which firstly relies on detailed information transferred from gem5. 
While existing gem5-to-McPAT connections rely on static mapping mechanisms and lack flexibility for integrating customized modules, our enhanced framework overcomes this limitation by providing a streamlined, user-configurable approach based on the features mentioned in \ref{tool_fusion_sec}. 
Architects can define translation rules for customized modules using a simple YAML configuration file. A parser script then automatically extracts and translates parameters from gem5 outputs based on the user-defined rules. This approach seamlessly integrates the customized modules into the McPAT power estimation flow without any modification to the parser code.

However, while fully customized module information is successfully passed from gem5 to McPAT’s input, McPAT itself is unable to process unknown modules that are not pre-programmed into its source code. 
Thus, an extension interface is added in McPAT to enable power simulation for customized modules. Customized modules are modeled at the block level, where power consumption could be calculated from static power, switching energy cost, activity factor, and switching pattern. In addition, the interface remains open for models with more details, as all computation-related functions are weak-attributed and can be freely overwritten. The extension interface composes of an XML input template and a power calculation template in the computation process. Consistent with McPAT’s original design, both physical parameters for model initialization and dynamic statistics for access patterns calculations are included in a unified XML file. For customized power simulation, all calculations occur inside McPAT alongside other system components, ensuring that system-level statistics are shared and interactions between customized modules and conventional units are accurately modeled. This integrated approach yields more reliable power estimation results compared to merging isolated power traces generated from different flows.


\section{Experiments and Results}

\begin{figure*}
    \centering
    \includegraphics[width=0.9\linewidth]{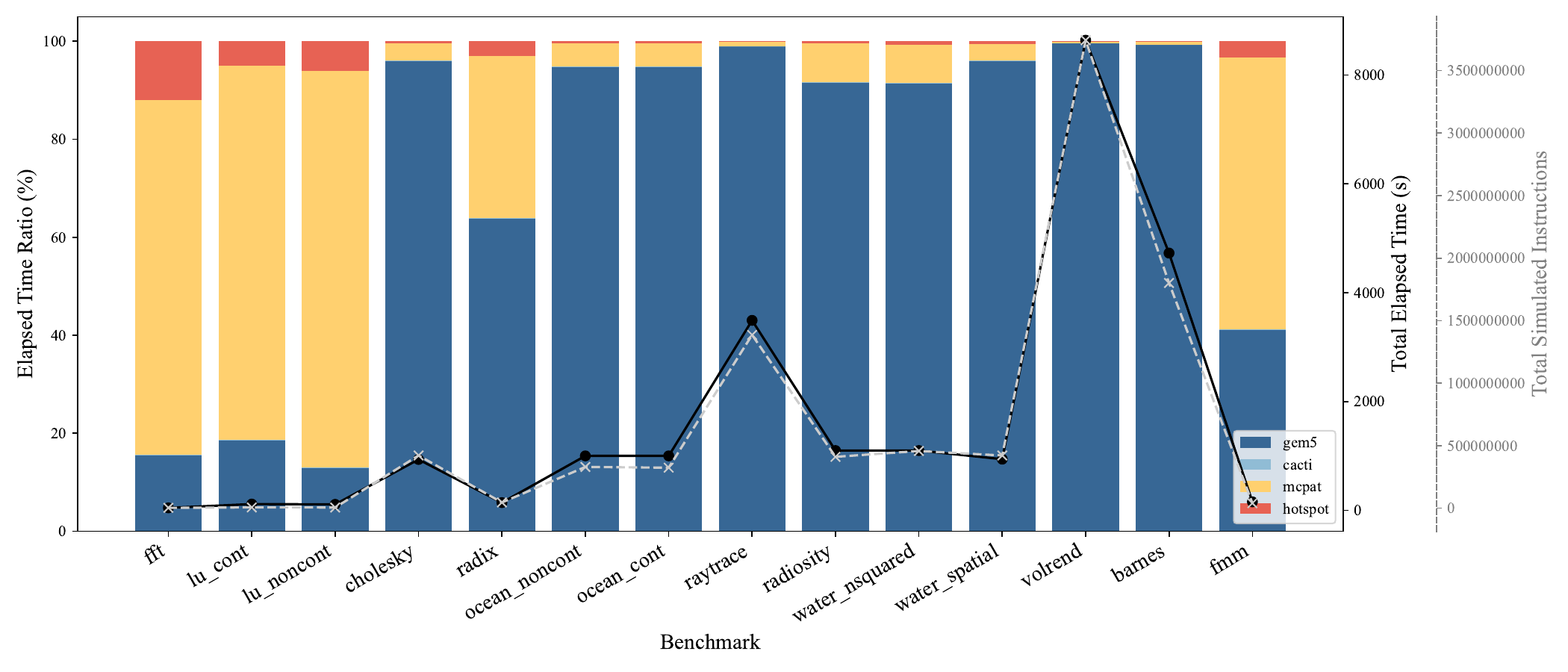}
    \caption{Elapsed time (black solid lines) and its breakdown (colored bars) for performance, power, and thermal models across the Splash2 benchmark suite with default problem size. Total simulated instruction counts (grey dashed lines) mark the correlation between sizes of workloads and total elapsed time}
    \label{fig:baseline_runtime}
\end{figure*}

To validate the effectiveness of the proposed Cool-3D framework and its key features, we conduct three case studies and present the corresponding validation results in this section. A detailed analysis will be performed to show how the thermal behaviors change along with changes in the design options. 


\subsection{Experimental Setup}
To better illustrate the features of Cool-3D and ensure fair comparisons among different cases, we construct a baseline 3D chip design that serves as the reference for improvements throughout the case studies. The baseline chip configuration is listed in Table \ref{tab:baseline_config}. The configuration of the baseline chip architecture is close to the setup of modern consumer processors. Additionally, the 3D-related configuration is set based on the existence of the combination of core and memory layers, and also the fact that those existing core-memory-stacked products have only a few layers. But Cool-3D does support modeling more layers and interleaving core and memory layers for next-gen processor or system-on-chip (SoC) development. The core model, along with the cache and memory settings, is configured through the Cool-3D front end and subsequently passed to the gem5 configuration interface. Following our tool integration workflow, McPAT and CACTI-3DD configurations are dynamically retrieved and incorporated during runtime after gem5 completes the architectural modeling. For 3D-specific parameters, in addition to those set in CACTI-3DD, stacking and cooling parameters are provided to HotSpot 7.0 via the Cool-3D front end at the start of the simulation flow. To evaluate the workloads run on the constructed hardware, we use the Splash2 benchmark suite \cite{splash2_artical, splash2_repo}, which is a commonly used testing program suite for computer architecture design performance evaluation, and run all benchmarks with their default problem size. 


\begin{table}[tb]
    \renewcommand{\arraystretch}{1.5}
    \caption{Configurations of the baseline chip}
    \label{tab:baseline_config}
    \centering
    \begin{tabular}{|c|c|}
    \hline
     Model Parameters & Values \\\hline
     Core Model & 4 cores, 2GHz, out-of-order, 8-issue \\\hline
     L1 I/D Cache & 2-way set associative, 4KB/8KB \\\hline
     L2 Cache & shared L2 cache, 2MB \\\hline
     Memory & 8GB, 32 banks, 8 banks per rank   \\\hline
    \end{tabular}\\\vspace{0.3cm}
    \begin{tabular}{|c|c|}
    \hline
        3D Related Parameters & Values \\\hline
        \multirow{4}{*}{Stacking} & 3 layers  \\\cline{2-2}
            & \textit{top layer} memory banks 16-31 ($4\times 4$) \\
            & \textit{middle layer} memory banks 0-15 ($4\times 4$) \\
            & \textit{bottom layer} cores 0-3 ($2\times 2$)    \\\hline
        Cooling & passive heat sink, no microfluidic cooling \\\hline
    \end{tabular}
\end{table}

\begin{figure*}[tb]
    \centering
    \includegraphics[width=0.9\linewidth]{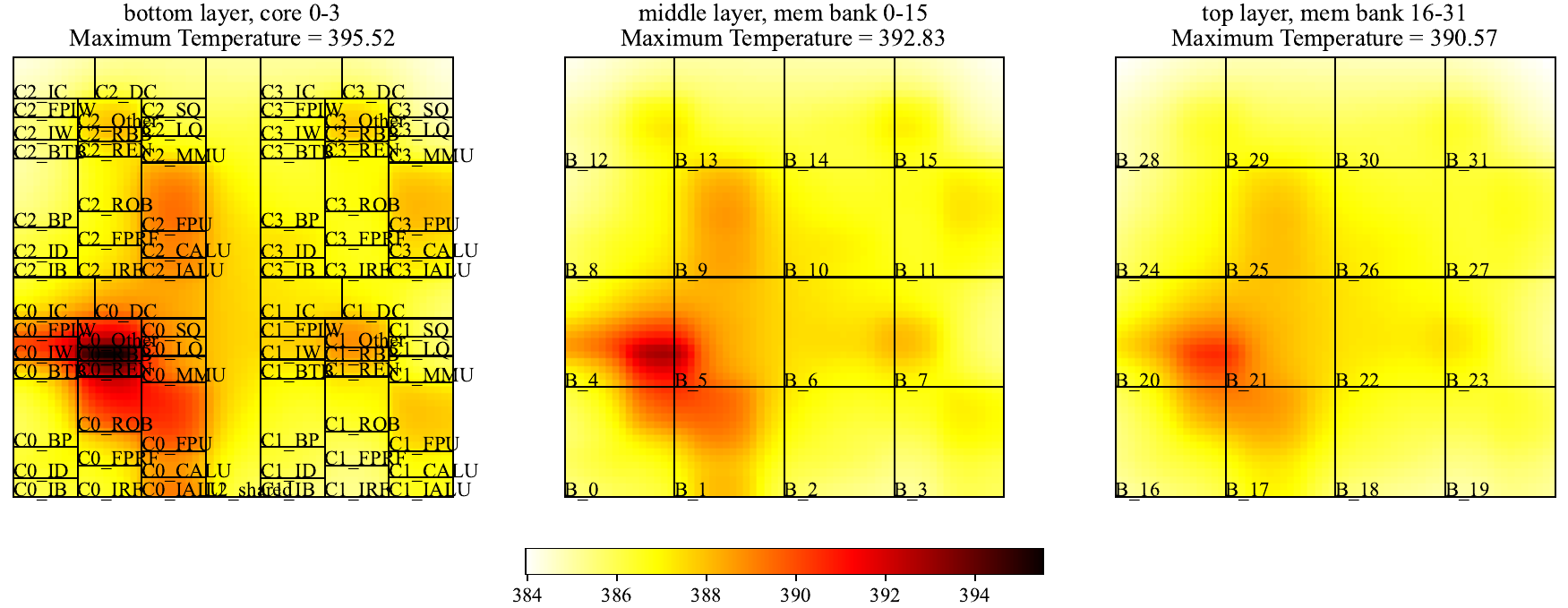}
    \caption{Heat maps (Temperature: K) of Splash2-fft benchmark running on the baseline chip. \textit{left}: core layer, \textit{middle}: memory layer 1, \textit{right}: memory layer 2.}
    \label{fig:baseline_only}
\end{figure*}

\begin{figure*}[tb]
    \centering
    \includegraphics[width=\linewidth]{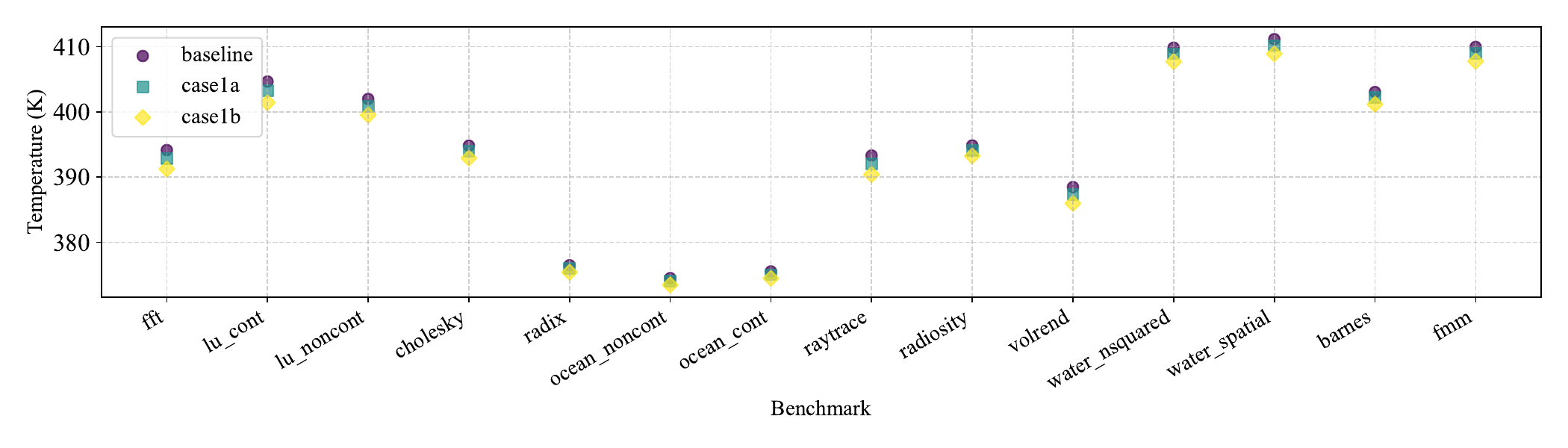}
    \caption{Temperature distribution for three cases: \textit{baseline}: core layer - memory layer - memory layer, \textit{case1a}: memory layer - core layer - memory layer, \textit{case1b}: memory layer - memory layer - core layer, across the Splash2 benchmark suite.}
    \label{fig:baseline_case1}
\end{figure*}

\begin{figure*}[htbp]
    \centering
    \includegraphics[width=\linewidth]{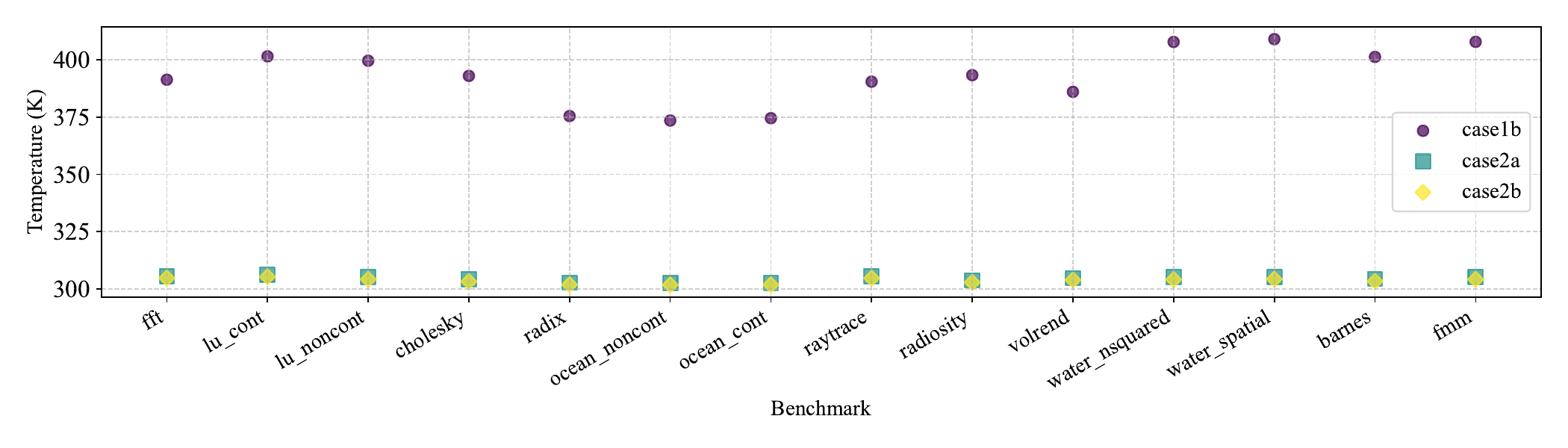}
    \caption{Temperature distribution for three cases: \textit{case1b}: core layer in the top, \textit{case2a}: microfluidic cooling with vertically aligned pattern applied, \textit{case2b}: microfluidic cooling with 90-degree bent microchannels applied.}
    \label{fig:case1_case2}
\end{figure*}

\begin{figure*}[htbp]
    \centering
    \includegraphics[width=0.9\linewidth]{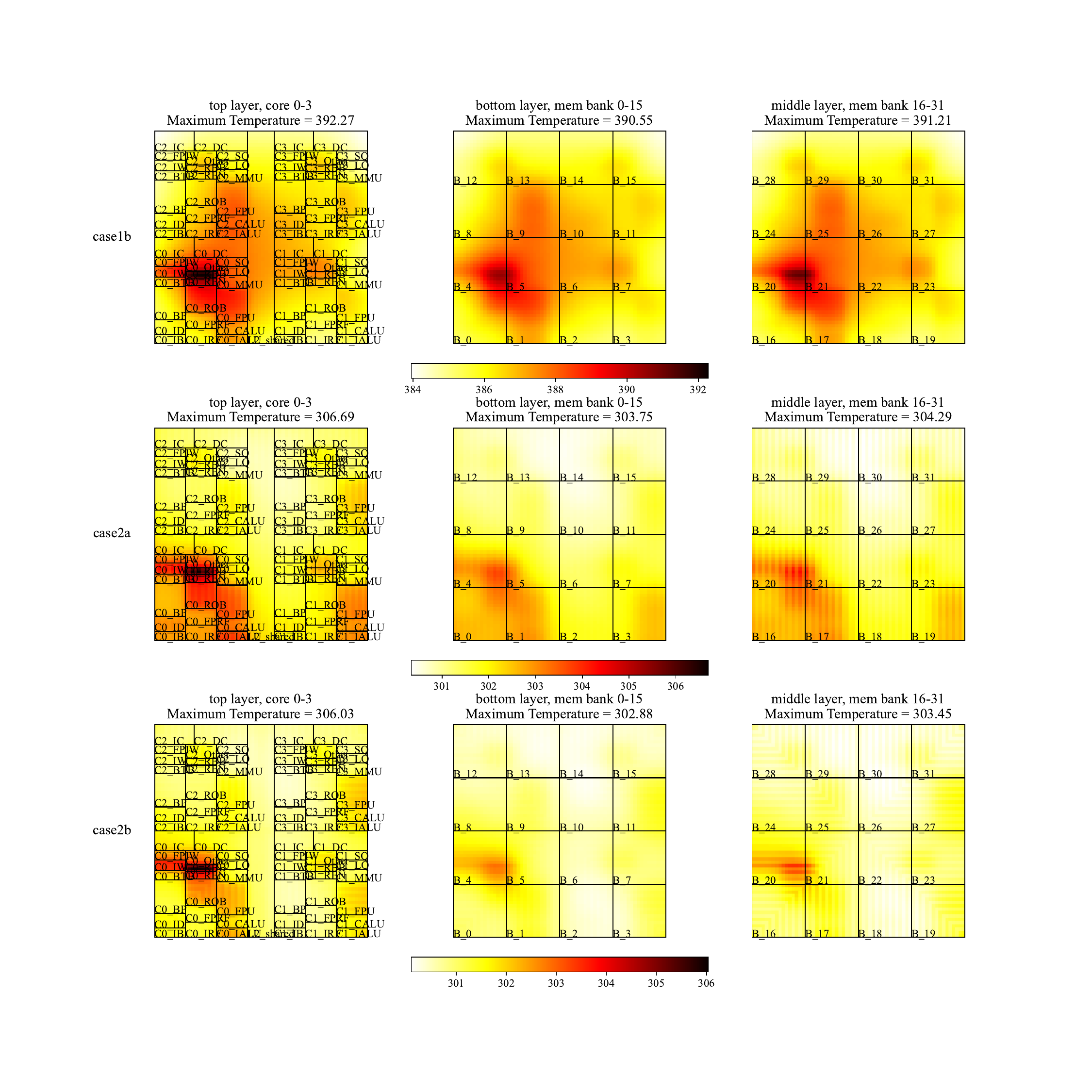}
    \caption{Heat map (Temperature: K) comparisons for \textit{case1b} (core layer in the top), \textit{case2a} (core layer in the top with microfluidic cooling with vertically aligned pattern), and \textit{case2b} (core layer in the top with microfluidic cooling with 90-degree bent microchannel pattern). }
    \label{fig:heatmap_comparison}
\end{figure*}

\begin{figure*}[htbp]
    \centering
    \includegraphics[width=\linewidth]{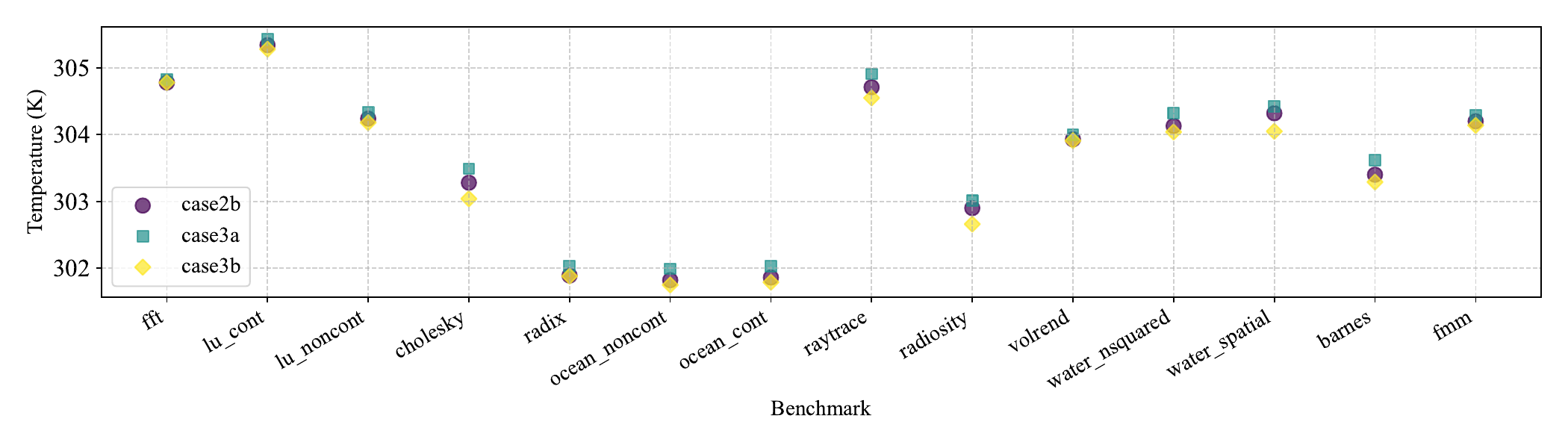}
    \caption{Temperature distribution for three cases: \textit{case2b}: core layer in the top with 90-degree bent microchannel styled microfluidic cooling applied, \textit{case3a}: shared L2 cache capacity scaled to 4MB based on \textit{case2b}, \textit{case3b}: shared L2 cache capacity scaled to 1MB based on \textit{case2b}.}
    \label{fig:case2_case3}
\end{figure*}

\subsection{Baseline Analysis}
The first experiment evaluates the baseline chip without any active cooling techniques. The entire process runs fully automatically without requiring additional user input. Upon completion, the experiment generates performance, power, and thermal results, along with a clear heatmap visualization to show the thermal distribution. 

Fig. \ref{fig:baseline_runtime} presents the elapsed time for running each benchmark, with a detailed breakdown of the time spent in each modeling phase. Due to the nature of performance modeling, the total runtime is highly workload-dependent. Additionally, Fig. \ref{fig:baseline_runtime} also shows the total number of simulated instructions in gem5 for each benchmark, where a clear correlation between instruction count and total elapsed time can be observed. On average, the end-to-end runtime for the entire flow across all benchmarks in default problem size is approximately 28 minutes, with the longest runtime reaching 2.4 hours. This efficiency allows designers to traverse the complete DSE flow with a complete benchmark suite within a few hours, enabling a rapid evaluation of potential design choices and their impact.

To further analyze thermal behaviors, we present the heat maps generated by running the benchmark \textit{fft} on the baseline chip. In Fig. \ref{fig:baseline_only}, three heat maps corresponding to the core layer and two memory bank layers are shown. These are obtained after running a steady-state simulation in the thermal model as the final step in Cool-3D. As shown in Fig. \ref{fig:baseline_only}, the core layer exhibits the highest temperature due to its greater power consumption and its position at the bottom of the stack. A closer examination of the core layer heat map reveals that core0 (indicated as ``C\_0'' in Fig. \ref{fig:baseline_only}) accumulates the most heat compared with others. This phenomenon aligns with the fact that in the performance trace, instructions executed on core0, on average, are $5.87\times$ more than the rest of the cores. From a microarchitectural perspective, the heat map helps pinpoint specific hotspots, such as floating-point computing units and result broadcast buses, which typically consume more power according to the power modeling results. Additionally, we observe coupling effects between adjacent layers, as indicated by the similar heat map patterns in the core layer and the adjacent memory layer 1.


\subsection{Case Study I - Altering Stacking Orders}
The baseline analysis has demonstrated that the position of the core layer within the stack significantly impacts heat dissipation, as accumulated heat is difficult to release from the backside. Consequently, stacking strategy plays a crucial role in thermal management. The most immediate solution to potential overheating issues in 3DIC design is to alter the stacking order. While this is straightforward for a two-layer stack, it becomes increasingly complex when multiple layers coexist, especially in the early design phase where the design is not even ready. 

In \textit{case1a}, we reposition the core layer between the two memory bank layers, whereas in \textit{case1b}, the core layer is placed at the top of the stack. Apart from the stacking order adjustments, all other parameters from Table \ref{tab:baseline_config} remain unchanged. To comprehensively evaluate the impact of stack ordering on overall temperature, we run Cool-3D with all benchmarks in both configurations. We use the maximum temperature of the entire 3D stack as the key comparison metric and optimization target, as it directly influences the final thermal design power (TDP)—a critical concern for architects and designers. The temperature variations across all benchmarks are illustrated in Fig. \ref{fig:baseline_case1}, where it is evident that \textit{case1b}, with the core layer positioned at the top of the stack, achieves the most significant temperature reduction across all workloads. This case study focuses on a three-layer stack, but Cool-3D is capable of configuring any number of layers, providing a flexible and accurate framework for 3DIC designers to conduct ``what-if'' analyses. By enabling rapid exploration of various stacking strategies, Cool-3D helps designers assess thermal implications and optimize layer placement for improved heat dissipation.



\subsection{Case Study II - Applying Microfluidic Cooling}
With Cool-3D, designers can now comprehensively evaluate the entire system while considering advanced cooling strategies, such as microfluidic cooling. This capability enables more accurate design budget estimations and facilitates potential cooling-architecture co-design. To demonstrate this, we update the baseline design to \textit{case1b} and use it as the new reference point for this round of optimization. In this case study, we compare two distinct microfluidic cooling patterns to determine the most effective heat dissipation strategy. 
In \textit{case2a}, the microchannels are arranged in a vertical configuration, as the previously discussed pattern in Fig. \ref{fig:cooling_pattern} (a), whereas in \textit{case2b}, the cooling pattern features two coolant inlets on the north and south sides and two outlets on the left and right, following the design outlined in Fig. \ref{fig:cooling_pattern} (b). 
It is important to note that the microfluidic cooling layer is placed between the top core layer and the middle memory bank layer, as the core layer continues to exhibit the highest temperature among all layers.

The temperature variations after applying the two different microfluidic cooling patterns are illustrated in Fig. \ref{fig:case1_case2}. A significant temperature reduction is observed across all benchmarks, demonstrating the effectiveness of microfluidic cooling. However, a noticeable difference exists between \textit{case2a} and \textit{case2b}, with \textit{case2b} featuring two coolant inlets and two outlets—achieving superior cooling performance compared to the vertically aligned pattern. To further analyze the impact of different cooling strategies, Fig. \ref{fig:heatmap_comparison} presents a heat map comparison for \textit{case1b}, \textit{case2a}, and \textit{case2b} using the \textit{fft} benchmark. The first row of Fig. \ref{fig:heatmap_comparison} displays the heat maps for \textit{case1b}, where the core layer is positioned at the top. The second row corresponds to \textit{case2a}, where the vertical microfluidic cooling pattern shown in Fig. \ref{fig:cooling_pattern} (a) is applied beneath the top core layer, maintaining the stacking order of \textit{case1b}. The last row represents \textit{case2b}, which applies the cooling pattern depicted in Fig. \ref{fig:cooling_pattern} (b). From this comparison, it is evident that, given the same stacking order, the cooling pattern illustrated in Fig. \ref{fig:cooling_pattern}(b) offers more effective heat dissipation, further validating the importance of optimizing microfluidic cooling strategies in 3DIC design.


\subsection{Case Study III - Applying Microarchitectural Changes}
To further demonstrate Cool-3D’s ability to capture subtle thermal variations induced by microarchitectural modifications in 3DICs, we introduce an adjustment to the shared L2 cache. Following the approach of the previous case study, we update the baseline design to \textit{case2b}, as it provides the best heat dissipation performance. Building on this new baseline, we modify the shared L2 cache capacity from its original 2MB to 4MB in \textit{case3a} and reduce it to 1MB in \textit{case3b}. The resulting temperature distributions are shown in Fig. \ref{fig:case2_case3}. The impact on temperature varies across benchmarks due to differences in data access patterns. However, in general, \textit{case3a} leads to a slight temperature increase across all benchmarks, whereas \textit{case3b} tends to reduce the temperature. This case study highlights the value of Cool-3D in identifying potential thermal issues arising from specific microarchitectural modifications, enabling designers to make informed decisions in optimizing 3DIC systems.


\subsection{Discussions}
Through Case Studies I–III, we emulate the design thinking process that a 3DIC designer might follow during the early design phase. While the actual design process is far more complex, involving a vast design space with numerous design parameters, especially when only conceptual designs are available. With Cool-3D, these design knobs can be easily configured to address high-level design questions, enabling key metric improvements such as better thermal profiles or, alternatively, more aggressive power budgets if sufficient cooling is guaranteed. While Cool-3D is not a one-size-fits-all signoff-calibre solution, its extension interface and flexible configurations offer a comprehensive set of design choices that are often scattered or overlooked in the early stages of 3DIC design.

\section{Conclusions and Future Work}
In this work, we introduce Cool-3D, an end-to-end thermal-aware DSE framework designed for fine-grained early-phase design space exploration of 3DICs. Cool-3D integrates a highly cohesive simulation flow with fine-grained configurability, enabling designers to explore a wide range of design options efficiently. The inclusion of microfluidic cooling modeling expands the design space, allowing for advanced thermal management strategies. Additionally, the built-in floorplan designer and microfluidic cooling strategy generator streamline configuration efforts. For non-parameterizable customizations, we provide an interface for integrating user-defined modules. Through extensive experiments, we demonstrate how Cool-3D effectively captures thermal-aware design variations across microarchitectural, stacking, and cooling dimensions, guiding designers toward more informed and optimized design decisions.


For future work, the first step would be to close the iterative DSE loop with novel optimization algorithms, such as reinforcement learning (RL) based methods, so that automatic tuning of the design configurations can be performed based on the thermal profile.  This enhancement will allow the framework to directly generate optimal designs according to designer-defined criteria. 
In addition, to broaden applicability and accommodate diverse simulator preferences, we plan to develop a ``super'' framework based on Cool-3D by incorporating alternative models. The ultimate goal is to establish a cross-validation thermal-aware framework that ensures both efficiency and accuracy in early-phase predictions. In the architectural design scope, future work will enhance support for emerging architectures, such as processing-in-memory (PIM) and processing-near-memory (PNM), leveraging the existing extension interface. This will enable more flexible and comprehensive thermal-aware design space exploration for emerging computing paradigms in 3DIC. The source code of this work is in the process of being released through \url{https://github.com/iCAS-SJTU/Cool-3D}. 


%

\section{Acknowledgment}
The authors would like to thank the developers and active contributors of Gem5, Sniper, CACTI, McPAT, HotSpot, SPLASH-2, and COMET for providing the foundational infrastructures that enabled this research. We also extend our gratitude to Dr. Jun-Han Han from the University of Virginia (now at Meta) for his valuable insights and constructive feedback during the early stages of this project.
\bibliography{reference}
\bibliographystyle{ieeetr}


 



\vskip -2\baselineskip plus -2fil
\begin{IEEEbiography}[{\includegraphics[width=1in,height=1.25in,clip,keepaspectratio]{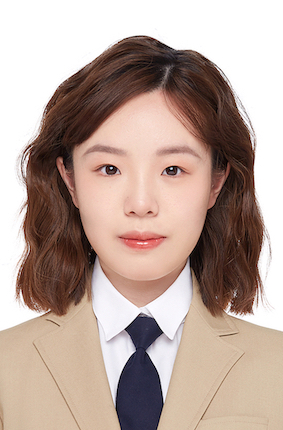}}]{Runxi Wang (Graduate Student Member, IEEE)}
is currently pursuing the Ph.D. degree at UM-SJTU Joint Institute at Shanghai Jiao Tong University, Shanghai, China. She received the B.E. degree in Electrical and Computer Engineering from Shanghai Jiao Tong University in 2023. Her current research interests include compute-in-memory and reconfigurable computing. She is currently the vice chair of the IEEE CASS SJTU Student Branch Chapter. 
\end{IEEEbiography}
\vskip -2\baselineskip plus -1fil
\begin{IEEEbiography}[{\includegraphics[width=1in,height=1.25in,clip,keepaspectratio]{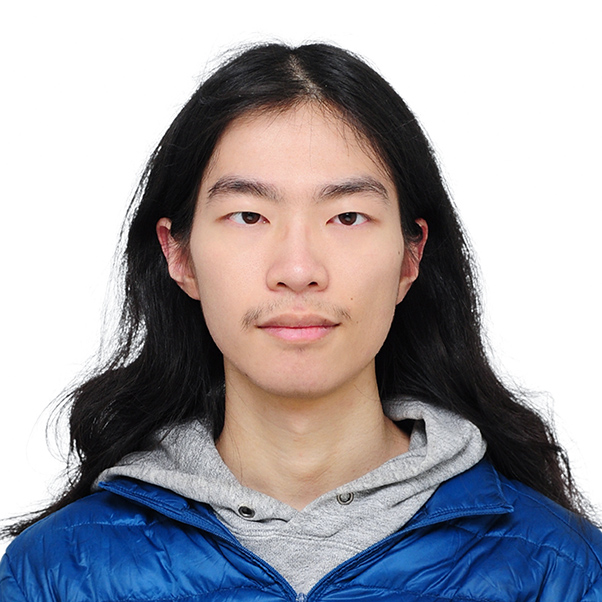}}]{Ziheng Wang}
is currently pursuing the B.E. degree at UM-SJTU Joint Institute at Shanghai Jiao Tong University, Shanghai, China. His current research interests include 3D ICs, computer architecture and computer system for AI.
\end{IEEEbiography}
\vskip -2\baselineskip plus -1fil
\begin{IEEEbiography}[{\includegraphics[width=1in,height=1.25in,clip,keepaspectratio]{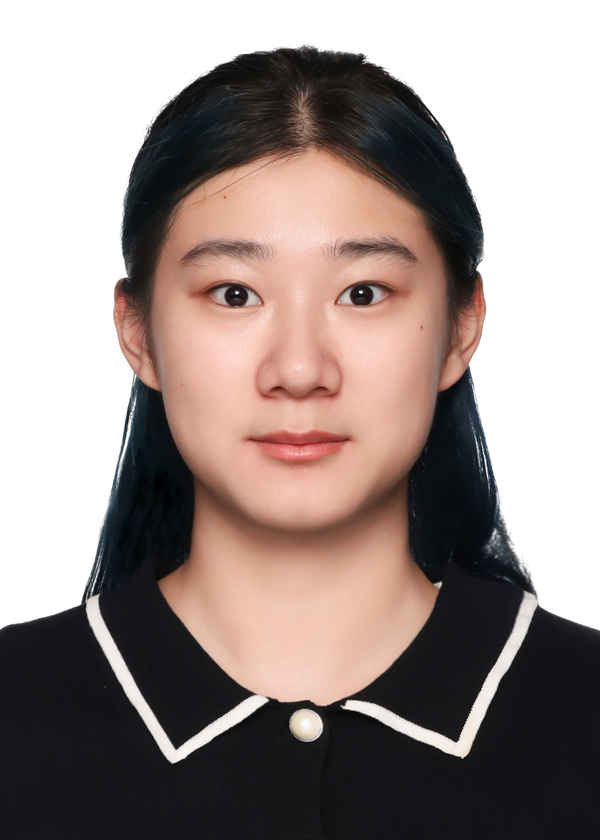}}]{Ting Lin (Student Member, IEEE)}
is currently pursuing the B.E. degree at UM-SJTU Joint Institute at Shanghai Jiao Tong University, Shanghai, China. Her current research interests include 3DICs and computer architecture for AI memory systems.
\end{IEEEbiography}
\vskip -2\baselineskip plus -1fil
\begin{IEEEbiography}[{\includegraphics[width=1in,height=1.25in,clip,keepaspectratio]{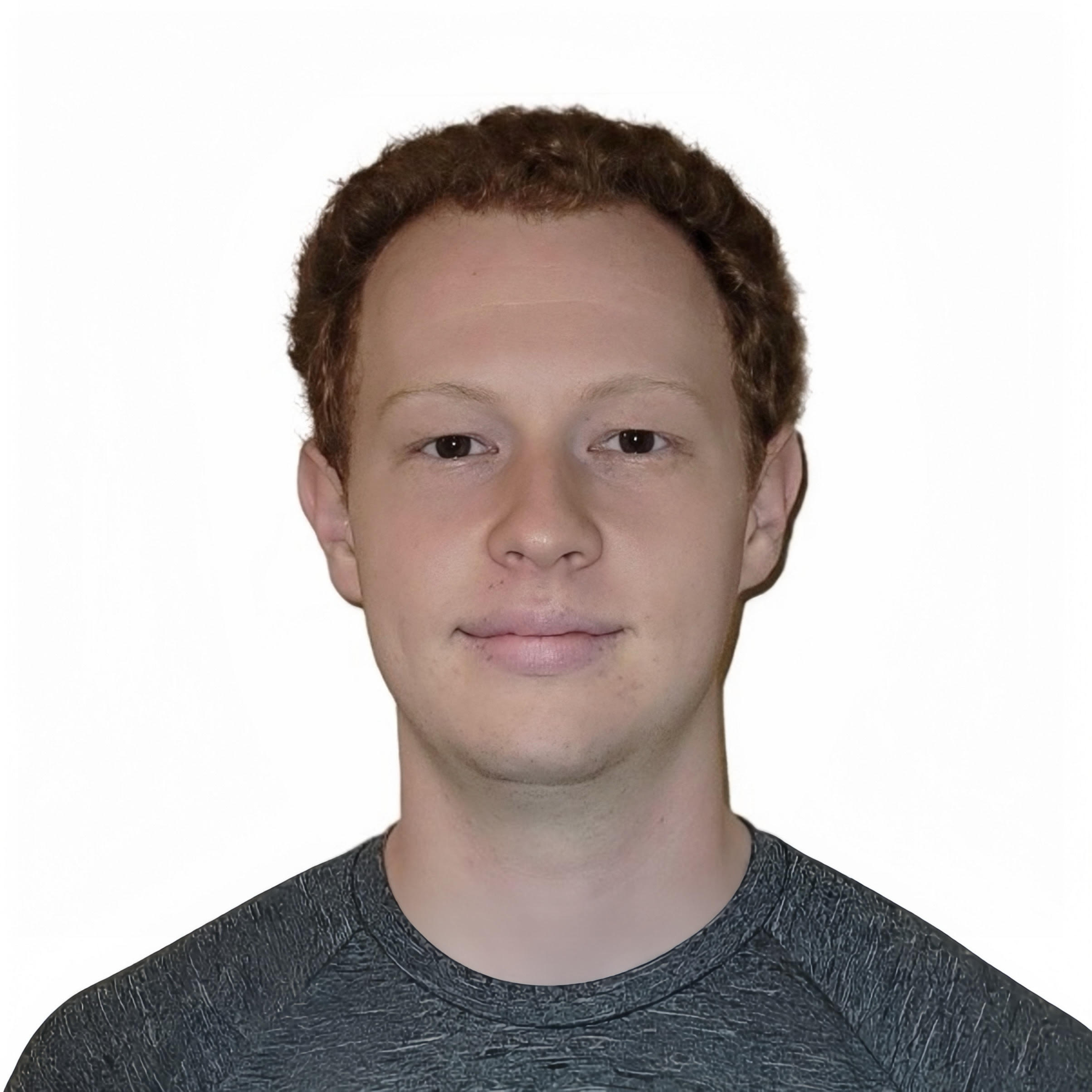}}]{Jacob Michael Raby (Graduate Student Member, IEEE)} is currently pursuing an M.S. degree in Electronic Science and Technology at the UM-SJTU Joint Institute, Shanghai Jiao Tong University, Shanghai, China. He received a B.S. degree in Computer Engineering from the University of Mississippi, Oxford, MS, USA, in 2024. His research interests include electronic design automation (EDA) tools, 3D ICs, and artificial intelligence.
\end{IEEEbiography}
\vskip -2\baselineskip plus -1fil
\begin{IEEEbiography}[{\includegraphics[width=1in,height=1.25in,clip,keepaspectratio]{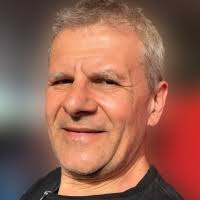}}]{Mircea R. Stan (Fellow, IEEE)}
is currently the Virginia Microelectronics Consortium Professor and Director of Computer Engineering at the University of Virginia, Charlottesville, VA, USA. Prof. Stan is teaching and doing research in the areas of AI hardware, Processing in Memory, Cyber-Physical Systems, Computational RFID, spintronics, and nanoelectronics. He is Editor-in-Chief for the IEEE TVLSI, and was a Distinguished Lecturer for the IEEE Circuits and Systems (CAS) Society in 2020-2021, 2012-2013 and 2004-2005, and for the Solid-State Circuits Society (SSCS) in 2007-2008.
\end{IEEEbiography}
\vskip -2\baselineskip plus -1fil
\begin{IEEEbiography}[{\includegraphics[width=1in,height=1.25in,clip,keepaspectratio]{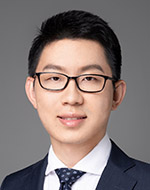}}]{Xinfei Guo (Senior Member, IEEE)}
is an Associate Professor with UM-SJTU Joint Institute at Shanghai Jiao Tong University (SJTU) in China. He received his Ph.D. in Computer Engineering from the University of Virginia. He serves as Associate Editor-in-Chief for IEEE TVLSI, Associate Editor for Integration, the VLSI Journal, and PC member or chair positions for over 30 international conferences. His research interests include software/hardware co-design for edge intelligence, machine learning-assisted EDA techniques and reconfigurable computing architectures. He is a senior member of ACM and IEEE.
\end{IEEEbiography}


\end{document}